\documentclass[preprint,showpacs,preprintnumbers,amsmath,amssymb]{revtex4}

\usepackage{graphicx}
\usepackage{dcolumn}
\usepackage{bm}
\usepackage{mathrsfs}

\begin{document}


\title{Cylindrical Lens by a Coordinate Transformation}

\author{Min Yan}
\author{Wei Yan}
\author{Min Qiu}
\email{min@kth.se}
\affiliation{%
Department of Microelectronics and Applied Physics, \\
Royal Institute of Technology, Electrum 229, 16440 Kista, Sweden
}%

\date{\today}

\begin{abstract}
Cylinder-shaped perfect lens deduced from the coordinate
transformation method is proposed. The previously reported perfect
slab lens is noticed to be a limiting form of the cylindrical lens
when the inner radius approaches infinity with respect to the lens
thickness. Connaturality between a cylindrical lens and a slab lens
is affirmed by comparing their eigenfield transfer functions.  We
numerically confirm the subwavelength focusing capability of such a
cylindrical lens with consideration of material imperfection. Compared
to a slab lens, a cylindrical lens has several advantages, including
finiteness in cross-section, and ability in lensing with magnification or
demagnification. Immediate applications of such a cylindrical lens can
be in high-resolution imaging and lithography technologies. In addition, its
invisibility property suggests that it may be valuable for non-invasive electromagnetic
probing.
\end{abstract}


\maketitle

The recent works on invisibility cloaking devices \cite{Leonhardt:conformal,Pendry:cloak} have
triggered a widespread interest on design of functioning
electromagnetic (EM) devices based on the coordinate transformation approach.
In an effort to unify a range of EM ``meta-phenomena'',
Leonhardt and Philbin point out that a perfect slab lens made of
negative index material (NIM) \cite{Pendry:perfectLens} can be interpreted as a result of a coordinate
transformation that maps a single region in virtual EM space to
multiple regions in physical space \cite{Ulf:Relativity}. A recent
extension of this concept for designing slab lenses with additional
operation functions, such as image translation, rotation,
and magnification etc, has been presented in Ref. \cite{Schurig:2007:slabLens}. In this
paper we report deployment of a spatial mapping in cylindrical coordinate for designing an annular
structure which can be considered as a cylindrical analog of the
previously much debated perfect slab lens. Such a cylindrical lens
overcomes two shortcomings of a slab lens. First,
its body has a finite cross-sectional size. No structural truncation
in cross-section is necessary for
its physical implementation\cite{note:finiteSize}. Second, it can form
magnified or demagnified image. These advantages can make such a lens
more favourable across many scenarios. It should be noticed that
Pendry has previously proposed a version of perfect cylindrical lens
which is based on rolling up a perfect slab lens existing in the virtual EM
space \cite{Pendry:OE-2003-clens}. However, some material parameters of Pendry's
cylindrical lens are gradient over the entire domain; in addition, the gradient parameters
approach infinity in the interior of the lens. Hence, a direct forward
theoretical confirmation of that lens is seemingly impossible, not to
mention the difficulty in its realization. In contrast, the
cylindrical lens under current study has all finite material
parameters, and an appropriate design can leave the exterior region
unaltered as free space. Therefore, the final device is much more
favourable from both realization and application points of
view.

\begin{figure}[h]
\centering
\includegraphics[width=8.5cm]{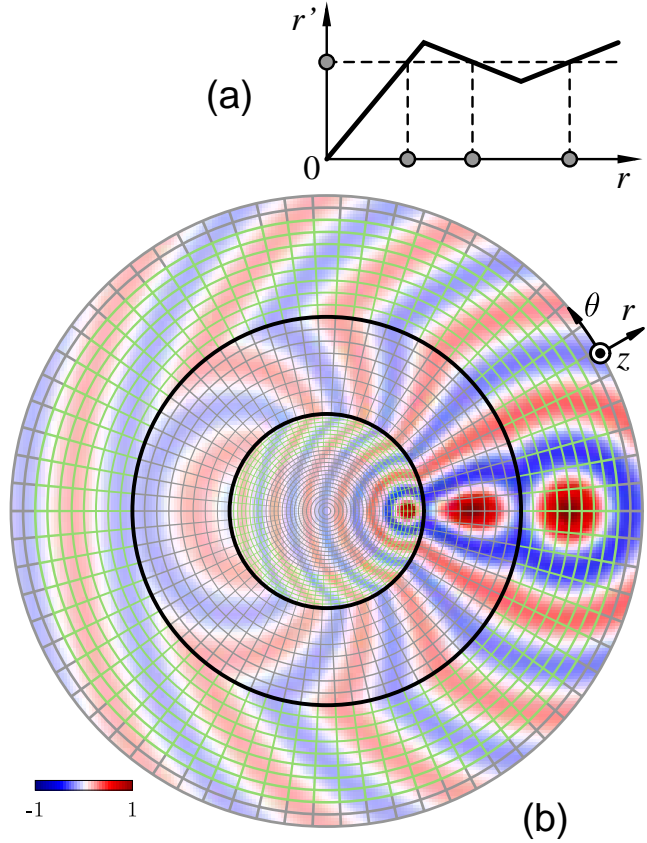} \\
\caption{Schematic illustration of the principle for idealized perfect
  cylindrical lensing. (a) Spatial mapping function. (b) Imaging of a
  line current source by a perfect cylindrical lens. Colormap shows the real
  part of the $E_z$ field.}
\label{fig:eventmap}
\end{figure}

The proposed cylindrical perfect lens is realized by taking a coordinate
transformation from a virtual free EM space ($r'$, $\theta'$, $z'$)
to the physical space ($r$, $\theta$, $z$), both denoted in
cylindrical coordinate. To achieve imaging in radial direction, we map
the coordinates as $r'= f(r)$, $\theta'=\theta$ and
$z'=z$.  One simple example of $f(r)$, as illustrated in
Fig. \ref{fig:eventmap}(a),  takes the form of
\begin{equation}
r'=
\begin{cases}
\frac{sa+(1-s)b}{a}r, &\text{if $r < a$;}\\
sr+(1-s)b, &\text{if $a \leq r \leq b$;}\\
r, &\text{if $r>b$,}
\end{cases}
\label{eq:map}
\end{equation}
where $a$ and $b$ are the inner and outer radii of the deduced
cylindrical lens (hence thickness $d=b-a$), and $s$ is the negative
slope deployed in the mapping function.  The continuity of the radial mapping function is
required in order to achieve
impedance-matched material interfaces. The negative slope at $a<r<b$
is responsible for transforming a single region in virtual space to multiple regions in physical
space. While the slope
can be of any negative value, in this paper we focus mainly on a slope of
$s=-1$. In Eq. \ref{eq:map}, the space outside the lens body is kept
unaltered, which allows the lens structure to work in a free space environment.

Coordinate transformation projects field in EM space directly to physical
space in accordance to the form-invariance property of the Maxwell equations. Consider a
line current source which is positioned at
$r'=\frac{3b-a}{2}$ and oriented along $z'$ axis in virtual space. Its
electric radiation field $E_{z'}'$ (a zeroth-order Hankel function) can be mapped
to the physical space via the relation $E_z(r)=E_{z'}'(r')$. The resulting
field distribution in physical space, illustrated in
Fig. $\ref{fig:eventmap}$(b) by its real part, shows as if there
are three current sources located at $r=\frac{3b-a}{2}$ (outside), $r=\frac{a+b}{2}$
(within), and $r=\frac{3ab-a^2}{4b-2a}$ (inside). The equivalence of
such multiple points in physical space implies that the device is able to
duplicate perfect copies of a source
\cite{Ulf:Relativity,note:eventMap}. The lens can image either in
inward mode (object is placed outside and its image is formed inside)
or reversely in outward mode with comparable performances.
Given the mapping function described by Eq. \ref{eq:map}, the object needs to be
placed within a certain distance to the lens' surfaces: outside
the lens, the effective region is $b<r_o\leq(2b-a)$; and inside $\frac{ab}{2b-a}\leq
r_o<a$.

The mapping function in Eq. \ref{eq:map} can be readily used to
understand the so-called hyperlenses proposed in Refs.
\cite{Tsang:PRB-2008-magnifyingLens,
  Kildishev:OL-2007-hyperlens,Kildishev:OL-2008-engineeringSpace}
under the context of cylindrical structure. These designs were
proposed to solve the impedance mismatch
problem of the multilayered cylindrical lenses theoretically analyzed in
\cite{Salandrini:PRB-2006-hyperlens, Jacob:OE-2006-hyperlens} and
demonstrated in \cite{Liu:Science-2007-hyperlens,
Smolyaninov:Science-2007-lens}. Under the framework of transformation
optics, the designs in \cite{Tsang:PRB-2008-magnifyingLens,
  Kildishev:OL-2007-hyperlens,Kildishev:OL-2008-engineeringSpace} are
inherently ``less perfect'' than the cylindrical lens proposed in this study. The
design method proposed in \cite{Tsang:PRB-2008-magnifyingLens} can be
interpreted as using an identical mapping function as in
Eq. \ref{eq:map}, but with a zero slope for $a<r<b$. For such a lens,
perfect imaging happens only when the object is placed exactly at $r=a$, and a
corresponding image at $r=b$. Notice also
that a coordinate transformation with a zero-sloped mapping
function leads to unphysical extreme material parameters. While in
\cite{Kildishev:OL-2007-hyperlens}, the design of a hyperlens can again be
interpreted as using Eq. \ref{eq:map} but with a close-to-zero positive
slope for $a<r<b$. No perfect imaging can be achieved in this
scenario. The lens relays an object's near field
at $r=a$ to $r=b$ with minimal
decaying, owing to the adjacency of $r=a$ and $r=b$ surfaces in
virtual space.

According to the coordinate transformation described by
Eq. \ref{eq:map}, we derive material
parameters for the designed cylindrical lens body ($a \leq r \leq b$) as
\begin{equation}
\epsilon_r=\mu_r=\frac{r-2b}{r},
\epsilon_\theta=\mu_\theta=\frac{r}{r-2b},
\epsilon_z=\mu_z=\frac{r-2b}{r},
\label{eq:MatParam}
\end{equation}
and for interior region $r<a$ as
\begin{equation}
\epsilon_r=\mu_r=1, \epsilon_\theta=\mu_\theta=1, \epsilon_z=\mu_z=\left(\frac{2b-a}{a}\right)^2.
\end{equation}
The lens has negative anisotropic permittivities and
permeabilities. At $r=b$, the lens has all material parameters valued at
$-1$, which is perfectly matched to air; at $r=a$, it has $\epsilon_r=\mu_r=\frac{a-2b}{a}$,
$\epsilon_\theta=\mu_\theta=\frac{a}{a-2b}$, and
$\epsilon_z=\mu_z=\frac{a-2b}{a}$, which matches to
the interior material. EM field in the interior region is compressed in wavelength
as compared to in free space due to its larger $\epsilon_z$ and
$\mu_z$ values. This contributes to image demagnification for inward
lensing operation, or magnification vice versa. It is worth mentioning that when the
lens thickness $d$ is very small compared to $a$, the lens' material
parameters at $r=a$ approach to $-1$, whereas $\epsilon_z$ and $\mu_z$
values of the interior material approach to $1$. That is, the
cylindrical lens physically becomes a previously reported prefect slab
lens \cite{Pendry:perfectLens}.

The perfect cylindrical lensing phenomenon predicted by coordinate
transformation, as illustrated in Fig. \ref{fig:eventmap}(b), however
should be verified using full-wave analyses. One factor for the
necessity of such explicit verification is that coordinate
transformation has presumed the time-harmonic steady state. In a previous
time-domain analysis for a slab lens, it was noticed that
re-construction of a  Fourier field component at
the other side of the lens can take impractically long time when
the slab index approaches to ideal value, i.e. -1 \cite{Rao:PRE-2003-transferFunction}.
This suggests that \emph{perfect} imaging by a slab lens is rather a
``wishful'' event which relies on infinite happening
time. Nevertheless, subwavelength imaging can be realized by slab lenses \emph{with
imperfect materials} \cite{Smith:APL-2003-limitation,Rao:PRE-2003-transferFunction}. Our
frequency-domain finite element calculation showed no convergence for
a cylindrical lens with ideal parameters as specified by
Eq. \ref{eq:MatParam}, which reflects the same singular problem of the
lens. In the following analyses, we examine
the performance of cylindrical lenses with the presence of material
perturbations.

For ease of clarification, we restrict our studies to in-plane
propagation case (wave vector in
$r\theta$ plane) with electric field solely polarized in
$z$ direction (E$_z$ polarization). It should however be kept in mind that coordinate
transformation promises focusing in a full 3D scenario with an arbitrary
wave polarization. The relevant material parameters hence
are $\epsilon_z$, $\mu_r$, and $\mu_\theta$. We append
a common factor $\varrho$, with $\varrho=1+\delta$, to each of the lens'
ideal material parameters. $\delta$ is referred to as material
perturbation, and its imaginary part represents material loss.
Material perturbation, especially loss, is inevitable for
resonance-based negative permittivity or permeability
metamaterials. For convenience, we denote layers in the
three-layered system from inside
to outside as layer 1, 2 and 3 respectively. General field solutions in
three layers, in terms of $E_z$, are
\begin{eqnarray}
E_{z}^1&=&\sum_m\left\{\mathscr{A}_m^1J_m\left(k_0\frac{2b-a}{a}r\right)+\mathscr{B}_m^1H^{(1)}_m\left(k_0\frac{2b-a}{a}r\right)\right\}\mbox{exp}(im\theta),
\label{eq:ez1}
\\
E_{z}^2&=&\sum_m\left\{\mathscr{A}_m^2J_m\left(k_0\varrho(r-2b)\right)+\mathscr{B}_m^2H^{(1)}_m\left(k_0\varrho(r-2b)\right)\right\}\mbox{exp}(im\theta),
\label{eq:ez2}
\\
E_{z}^3&=&\sum_m\left\{\mathscr{A}_m^3J_m(k_0r)+\mathscr{B}_m^3H^{(1)}_m(k_0r)\right\}\mbox{exp}(im\theta),
\label{eq:ez3}
\end{eqnarray}
where $k_0$ is free-space wave number, $m$ is the azimuthal momentum
number, and $\mathscr{A}^i$ and
$\mathscr{B}^i (i=1,2,3)$ are coefficients. We take $\mbox{exp}(-j\omega
  t)$ time harmonic dependence. Therefore the Hankel function of the
first kind $H^{(1)}_m$ in Eqs. \ref{eq:ez1}-\ref{eq:ez3} represents an outgoing
cylindrical wave.
Unlike in a planar system where purely evanescent or
propagating eigenwaves exist, all eigenwaves in the cylindrical lens
system are propagating. Despite the dissimilarity in eigenwaves, we
show in the following that superlensing mechanisms of two types
of lens can be bridged.

\begin{figure}[h]
\centering
\includegraphics[width=8.5cm]{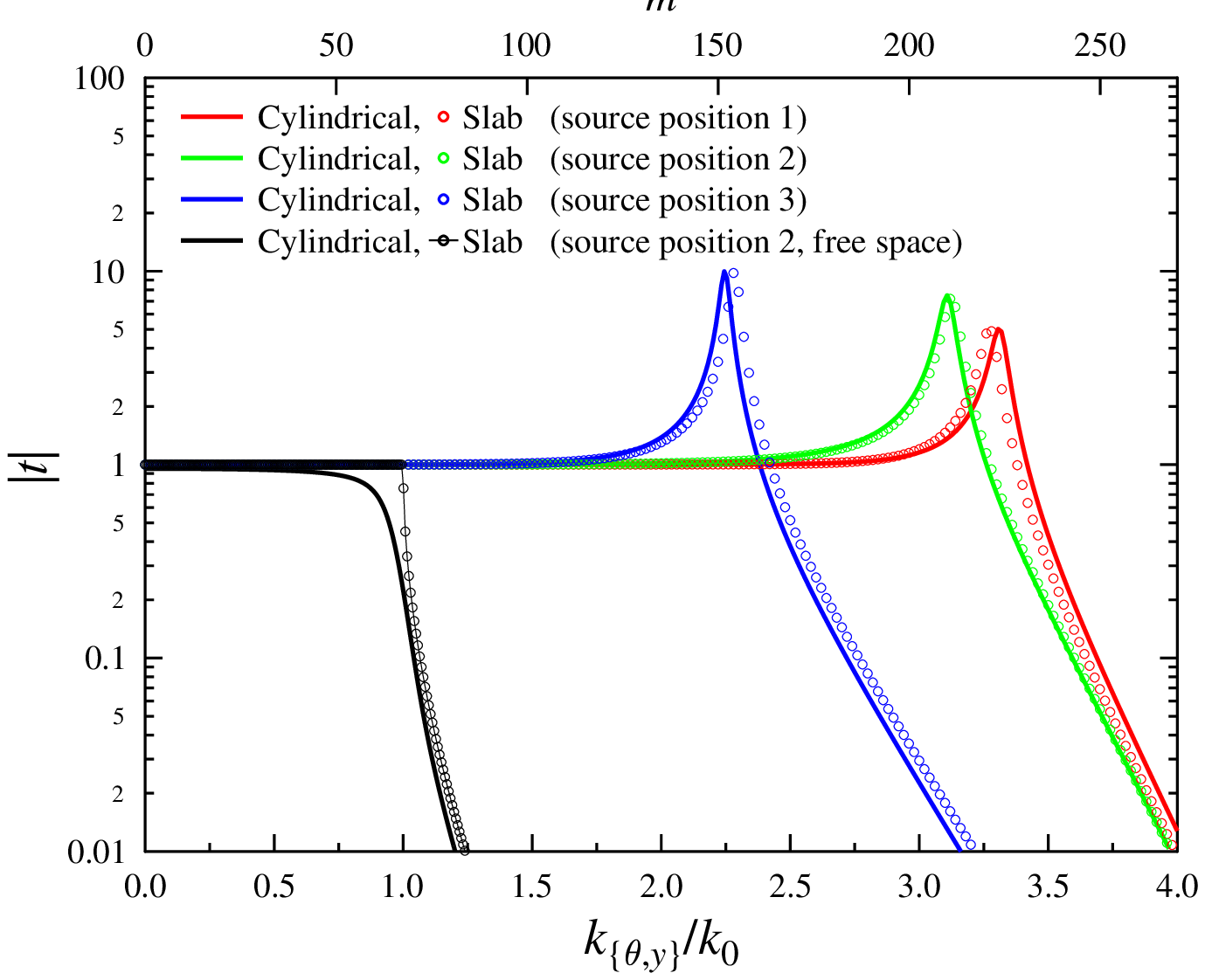} \\
\caption{Comparison of transfer functions for a cylindrical lens and a
  slab lens of the same thickness. Three different object positions
  are shown: for cylindrical lens $r_{o,1}=\frac{5ab-a^2}{8b-4a}$,
  $r_{o,2}=\frac{3ab-a^2}{4b-2a}$,
  $r_{o,3}=\frac{7ab-3a^2}{8b-4a}$; and for slab lens
  $x_{o,1}=-\frac{5d}{4}$, $x_{o,2}=-d$, $x_{o,3}=-\frac{3d}{4}$. Two
  sets of object positions are equivalent when $a$ and $b$ are much
  larger than $d$. For the cylindrical lens the corresponding angular
  momentum number is marked on the secondary $x$ axis.}
\label{fig:TFCylinSlab}
\end{figure}

First, we consider a cylindrical lens whose inner
radius $a$ is much greater than both its thickness $d$ and the operating
wavelength. The system
resembles a slab lens in both geometry and material parameters. Our objective
is to show how the imaging performances of such a cylindrical lens and
a slab lens of the same thickness can be quantified in a unified
manner. For a slab lens, the transfer
function, which describes the ratio of field
strengths at image and source positions for each
plane wave component, has been used for describing its imaging
capability \cite{Smith:APL-2003-limitation,
  Rao:PRE-2003-transferFunction, Podolskiy:OL-2005-lens}. Consider a
slab lens with thickness $d=\lambda/2$ which is placed in
$yz$ plane and centered at $x=0$. Likewise, we assume that waves are
propagating in $xy$ plane with pure E$_z$ polarization. The slab lens has
material parameters of $\epsilon=\mu=-\varrho=-1-\delta$, with $\delta=0.0001-0.00001i$.
Its transfer functions, as characterized by normalized $k_y$ are shown in Fig.
\ref{fig:TFCylinSlab} (solid curves) for three different object
positions. For this particular material perturbation introduced, restoration of
evanescent waves ($k_y>k_0$) at the image positions (always $2d$ apart
from the source) has been realized up to a certain
$k_y$. The transfer functions are characterized by a resonance
before the transmission damps heavily. The free-space transfer
function from the source position to image position (without the slab
lens) is also included in Fig. \ref{fig:TFCylinSlab}. The evanescent
waves can barely reach the image position. For a cylindrical lens, one may
first be puzzled by the absence of decomposable wavevectors in the
general solutions (Eqs. \ref{eq:ez1}-\ref{eq:ez3}). In fact, the
azimuthal momentum number $m$ determines the pace of field variation
along the surface-tangent ($\theta$) direction. The azimuthal momentum number can be converted to azimuthal
wave number as $k_\theta=m/r_i$, where $r_i$ is the radial image
position \cite{note:aziWV}. In order to obtain
sharp image especially in $\theta$ direction, existence of high-order
cylindrical wave components \emph{at the image position} is
essential. Hence, transmission of cylindrical waves through a
cylindrical lens provides vital information for the lens' imaging
capability. Here we consider a
cylindrical lens with $d=\lambda/2$ and $a=10\lambda$. $\delta$ is
kept the same as in the slab case. The cylindrical wave transfer functions for
outward lensing operation, plotted in terms of normalized
$k_\theta$, are superimposed in Fig. \ref{fig:TFCylinSlab}. Three
different object positions have been studied in
Fig. \ref{fig:TFCylinSlab}. The
transfer function records the ratio of field strengths at the image and
source positions at different cylindrical wave order number $m$. For
each $m$, the field ratio is derived through matching of the
tangential fields at both $r=a$ and $r=b$ interfaces. Referring to
Eqs. \ref{eq:ez1}-\ref{eq:ez3}, for outward imaging operation, the task
is to find the coefficient $\mathscr{B}^3$ from a known value of
$\mathscr{B}^1$; while for inward imaging operation, the task is to
find the coefficient $\mathscr{A}^1$ from a known value of
$\mathscr{A}^3$. Unlike for a slab lens case, the transfer function
for a cylindrical lens is too lengthy to be expressed here owing to
the presence of transcendental Bessel functions.
For consistency, the free-space cylindrical wave transfer function is
also shown. From Fig. \ref{fig:TFCylinSlab}, we notice excellent
agreement between the transfer functions for cylindrical and slab
lenses. With the same thickness and material perturbation, the two
types of lenses are therefore equivalent in their imaging
capabilities. Both of them can restore a spectrum of waves
($k_{\{\theta, y\}}>1$) that were lost in the absence of the
lenses. Figure \ref{fig:TFCylinSlab} provides independent
verification of the sub-wavelength imaging capability of a cylindrical
lens.

\begin{figure}[h]
\centering
\includegraphics[width=8.5cm]{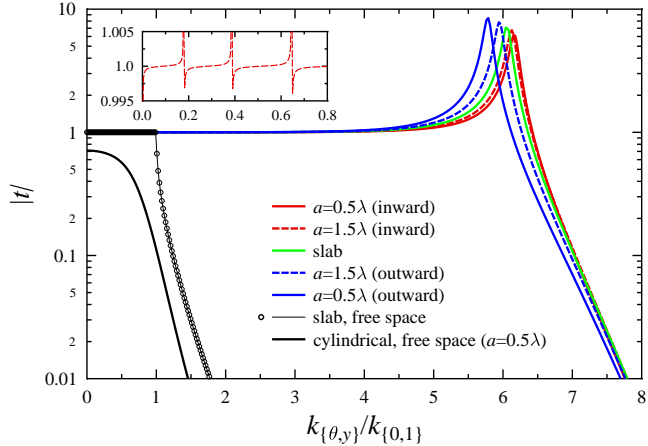}
\caption{Effect of $a$ on cylindrical wave transfer function. Inset
  gives the zoom-in view of the transfer function with $a=1.5\lambda$
  (inward).}
\label{fig:effect_a}
\end{figure}

Next, we look into the imaging performance of a cylindrical lens as
its inner radius $a$ decreases. At a smaller $a$ value, the physical
problem becomes more distinct as compared to a slab lens. It is worth
noting that, when the object and image points under examination are
very close to the origin of cylindrical system, a cylindrical wave
can appear to be a strong mixture of both propagating and evanescent
waves. This can be observed by comparing the free-space transfer
functions in slab and cylindrical systems shown in
Fig. \ref{fig:effect_a}. Having known this, one would
naturally expect to see the distinct character of a cylindrical lens
as reflected in its transfer function. The result
however turns out to be a bit surprising. The cylindrical wave
transfer function is hardly affected when its inner radius is reduced
even close to wavelength. In our case study, we fix thickness
$d=0.25\lambda$ and decrease $a$ from infinity to
$0.5\lambda$. Material perturbation is again held at $\delta=0.0001-0.00001i$.
Besides showing the effect of reducing $a$ on the transfer function, we also take this
opportunity to examine the difference in performances between two
operation modes of the same lens, i.e. inward imaging and outward
imaging. For each operation mode, we start from the slab lens case ($a=\infty$),
and decrease $a$ to  $1.5\lambda$ and then to $0.5\lambda$. The variation of
the transfer function is shown in Fig. \ref{fig:effect_a}. Notice that for inward operation,
$k_\theta$ is normalized with respect to the wavenumber in layer 1,
i.e. $k_1=k_0\frac{2b-a}{a}$. Object position for inward operation is
at $(3b-a)/2$, and for outward operation $(3ab-a^2)/(4b-2a)$. Figure
\ref{fig:effect_a} shows that the transfer function shifts only very
slightly when $a$ decreases from infinity to $0.5\lambda$. Therefore
it can be concluded that, with the same thickness and material
perturbation, a cylindrical lens, even at an overall scale comparable to wavelength, is
as effective as a slab lens for achieving subwavelength imaging. This
conclusion is valid for either inward or outward operation mode. The
sharp variations of the inward transfer functions at small
$k_\theta$ values (see the inset in
Fig. \ref{fig:effect_a}) are due to near zero field values at both
object and image positions. They are not caused by resonances.

\begin{figure}[h]
\centering
\includegraphics[width=8.5cm]{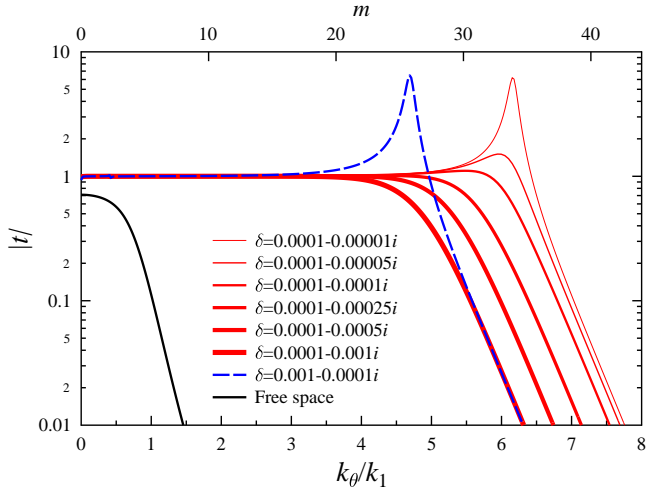}
\caption{Effect of $\delta$ on cylindrical wave transfer function. Azimuthal momentum number is marked
  on the secondary $x$ axis.}
\label{fig:effect_delta}
\end{figure}

The inevitable material
imperfectness imposes a rather stringent limitation on
the lens shell thickness. For a slab lens, it has been noticed that a smaller thickness
drastically improves the imaging performance (or equivalently, relaxes the requirement
for material perfectness in order to achieve the same subwavelength imaging)
\cite{Smith:APL-2003-limitation}. We have noticed the same for cylindrical lenses. This
information can be obtained by comparing Fig. \ref{fig:TFCylinSlab}
and Fig. \ref{fig:effect_a}. In general, with a reasonable material
perturbation, the thickness of a cylindrical lens
should be kept less than a wavelength in order
to achieve subwavelength resolution. Here we explicitly look into the
effect of material perturbation on a relatively thin
cylindrical superlens' performance. The cylindrical lens to be
studied has fixed geometrical parameters as
$a=0.5\lambda$ and $d=0.25\lambda$. First the effect of
material loss is examined by keeping the real part of $\delta$, or $\Re(\delta)$,
at $0.0001$, and increasing the imaginary part $\Im(\delta)$ from
$0.00001$ up to $0.001$. The variation of the transfer function is shown in
Fig. \ref{fig:effect_delta}. From the figure, it is
noticed that if $\Im(\delta)<\Re(\delta)$, only the ``finesse'' of the resonance peak
is affected. Higher loss suppresses the resonance.
Only when $\Im(\delta)>\Re(\delta)$,
the transmission starts to drop at a significantly lower $k_\theta$
value, which therefore affects the image resolution. A material perturbation
dominated by either its real part or imaginary part
(but with around the same value) will result in
similar transfer functions, with the exception that no resonance
peak is present for the transfer function resulted from the latter
perturbation. This is confirmed by two curves ($\delta=0.0001-0.001i$
and $\delta=0.001-0.0001i$) in Fig. \ref{fig:effect_delta}. The
imaging enhancement factor, defined as the maximum
$k_\theta/k_1$ at which the transmission from the object to image is
close to unity is about 4.5 for
$\delta=0.0001-0.001i$ and 6 for $\delta=0.0001-0.00001i$. The results
suggest that cylindrical lenses under study are able to image with
subwavelength resolution. When the perturbation is reduced further,
the resonance peak will shift to an even larger $k_\theta$ value. In the
limit, for an ideal lens the resonance happens at infinite $k_\theta$.

\begin{figure}[h]
\centering
\includegraphics[width=8.5cm]{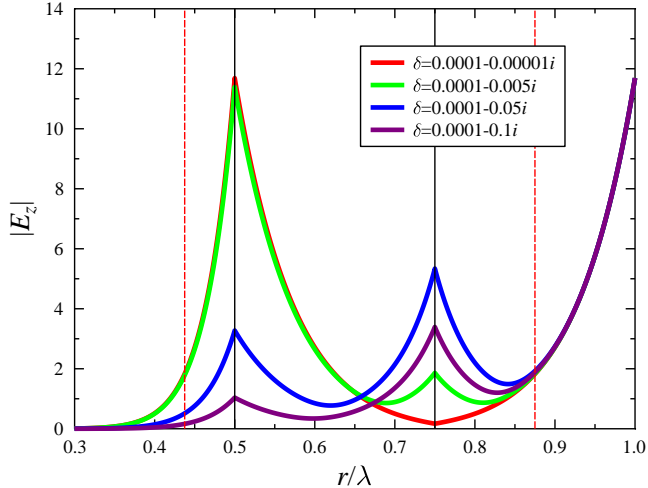}
\caption{Radial $|\mathrm{E}_z|$ field distributions as material loss increases. The
  vertical solid black lines denote lens boundaries; dashed red lines
  denote object and image positions.}
\label{fig:fld_trans}
\end{figure}

In Fig. \ref{fig:fld_trans}, we show the radial $|E_z|$ field
distribution of a high-order cylindrical wave ($m=15$) for the
lens studied in Fig. \ref{fig:effect_delta}. Particularly, we increase
the loss perturbation and observe its effect on field
profile. $\Re(\delta)$ is kept at 0.0001, and $\Im(\delta)$ is changed
from 0.00001 to 0.005, 0.05, and 0.1. The study is for inward imaging
scenario. The cylindrical wave, at the
vicinity of the object or image position, resembles a purely
evanescent wave. The field only becomes oscillating at a
distance further away from the origin. With a very small loss, the
incoming field from outside penetrates freely (through the formation of a
so-called anti surface mode at the outer interface) into the lens body and
excites a surface mode at the inner lens boundary. The field amplitude
is therefore amplified. With a further decay in lens interior, the
field amplitude at the image position is almost exactly the same as at
the object position. Restoration of the cylindrical wave is
realized. When the loss increases, a surface mode starts to build up at
the outer boundary of the lens, while the strength of the surface mode
at the inner boundary is gradually muted. At $\Im(\delta)=0.01$, the
loss becomes so heavy that the first surface mode also weakens. The
results shown here are very similar to those noticed for a slab lens \cite{Rao:PRE-2003-Amplification}.

\begin{figure}[h]
\centering
\includegraphics[width=6cm]{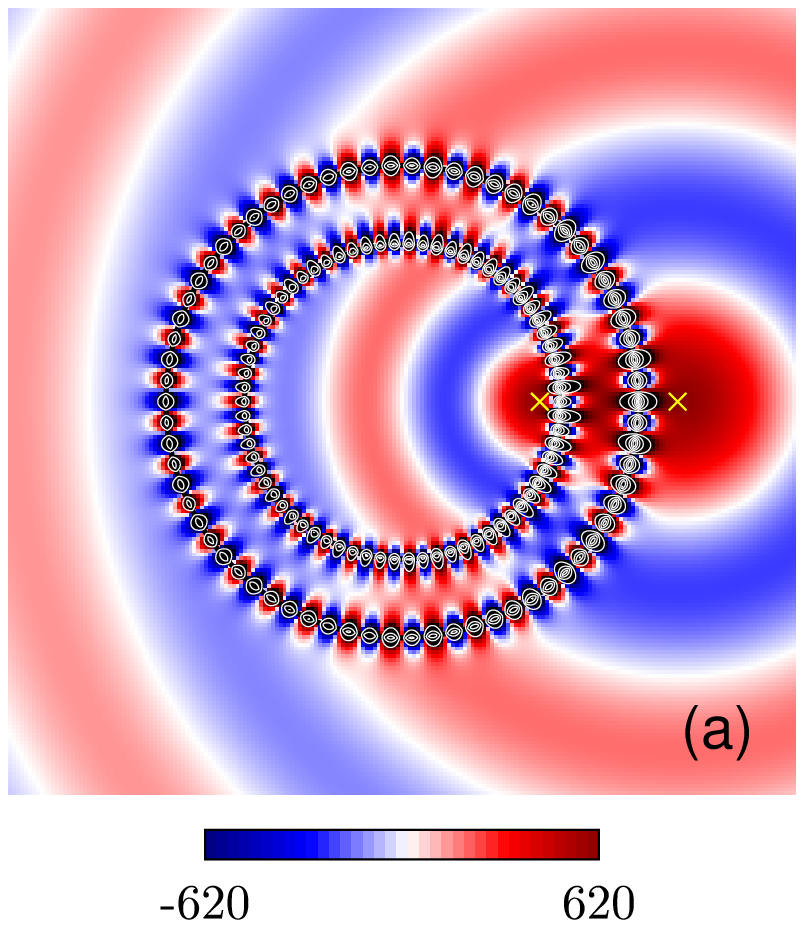}
\ \
\includegraphics[width=6cm]{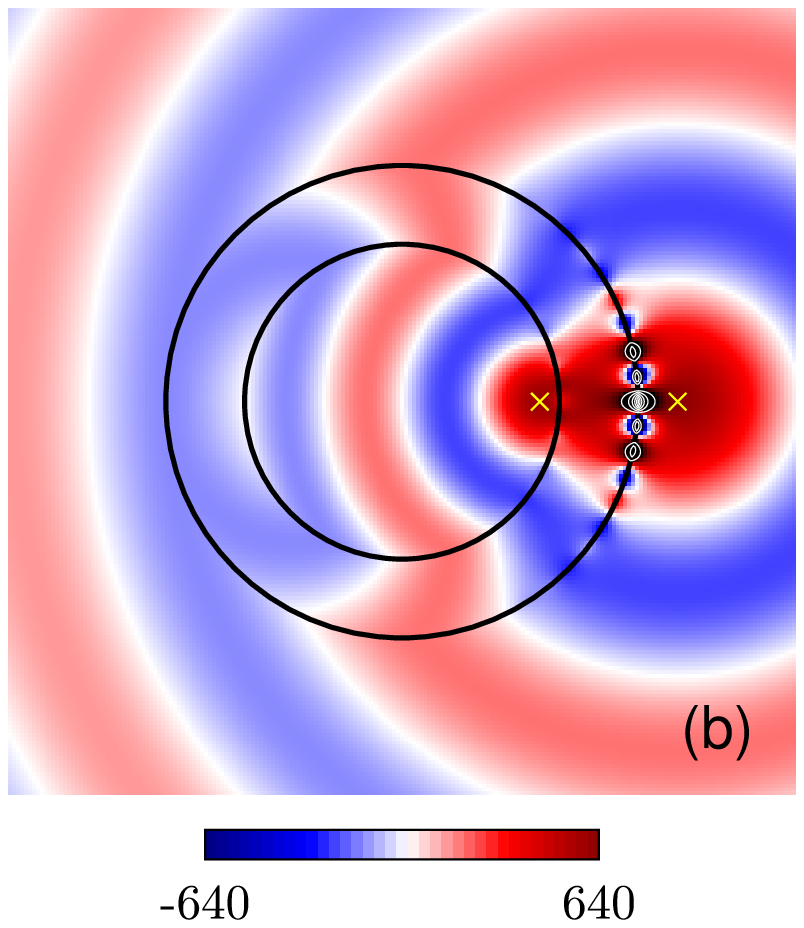}
\caption{Inward imaging of a line current source. (a)
  $\delta=0.0001-0.00001i$. Overvalued field
are shown in saturated color as dark red (positive) or dark blue
(negative), and contour lines at the
levels of $\pm 1000$, $\pm 2000$, $\pm 3000$, $\pm 4000$, $\pm 5000$
and $6000$ are marked. (b) $\delta=0.0001-0.001i$. Contour lines at the
levels of $\pm 1000$, $\pm 1500$, $2000$, $2500$, $3000$
and $3500$ are marked. Yellow crosses are object and image
positions. Domain size: $2.5\lambda\times 2.5\lambda$.}
\label{fig:src_1}
\end{figure}

To complement the above transfer function analyses, here we numerically
demonstrate the imaging performance of
cylindrical lenses using the finite element method (FEM). The issue of coupled
surface-mode resonance, which gives rise to the peaks in
the transfer functions, is also addressed here. The first lens to be
examined has parameters of $a=0.5\lambda$ and $d=0.25\lambda$,
and $\delta=0.0001-0.00001i$. We position a $z$-oriented line current source
at $r_o=(3b-a)/2$. An image is expected at $r_i=(3ab-a^2)/(4b-2a)$ in
lens interior. The real part of the steady state $|E_z|$ field is captured in
Fig. \ref{fig:src_1}(a). An image is clearly re-constructed inside the
lens. However, the most noticeable difference as compared to the
idealized field distribution in Fig. \ref{fig:eventmap} is that
strong localized fields are present at the two
surfaces of the lens. The maximum field values on the surfaces
are as high as 6500 positively and -5000 negatively (in
arbitrary unit), which are about 10 times as large as the value at the object
position. There are altogether 70 nodal lines along
azimuthal direction, indicating the resonance happens around
$m=35$, which agrees with its corresponding transfer function in
Fig. \ref{fig:TFCylinSlab}. The imaginary part of the field (not
shown) has the same
number of nodal lines, but with different positions as compared to the
real part. Hence the surface wave is not standing, but
traveling along $\theta$ direction. An examination of the Poynting
vector plot confirms giant energy flow around the lens'
surfaces (not shown). The huge amount of energy that
the lens can trap suggests that it may require a certain amount of
time for the system to reach its steady state, and in turn to form the
image. Another consequence of the existence of heavy surface modes is
that it may disallow an object from being placed very near or very
far to the lens surface, which adds further restriction on the
effective lensing and detecting areas. From
Fig. \ref{fig:effect_delta}, we know that a relatively lossy material
can help to suppress the resonance peak in transfer
function, though at the expense of a lower achievable resolution. We
therefore carried out another FEM analysis for the same
structure but with $\delta=0.0001-0.001i$. The real part of the $E_z$
field is shown in Fig. \ref{fig:src_1}(b). Compared to
Fig. \ref{fig:src_1}(a), complete suppression of the surface mode is seen
at surface positions away from the line source. Still, surface mode is present
close to the line source, but its strength has been reduced
considerably compared to the previous calculation. For both cases,
despite the localized surface modes, field \emph{outside the lens}
experiences almost no disturbance by the lens
structure; it appears to be radiated directly from the line
source. Therefore the lenses are almost invisible to an observer. The
invisibility is due to the fact that the particular coordinate
transformation deployed assures the whole structure bounded by $r<b$
is electromagnetically equivalent to vacuum. The perturbations
introduced here are insignificant for affecting the invisibility
property.

\begin{figure}[h]
\centering
\includegraphics[width=14cm]{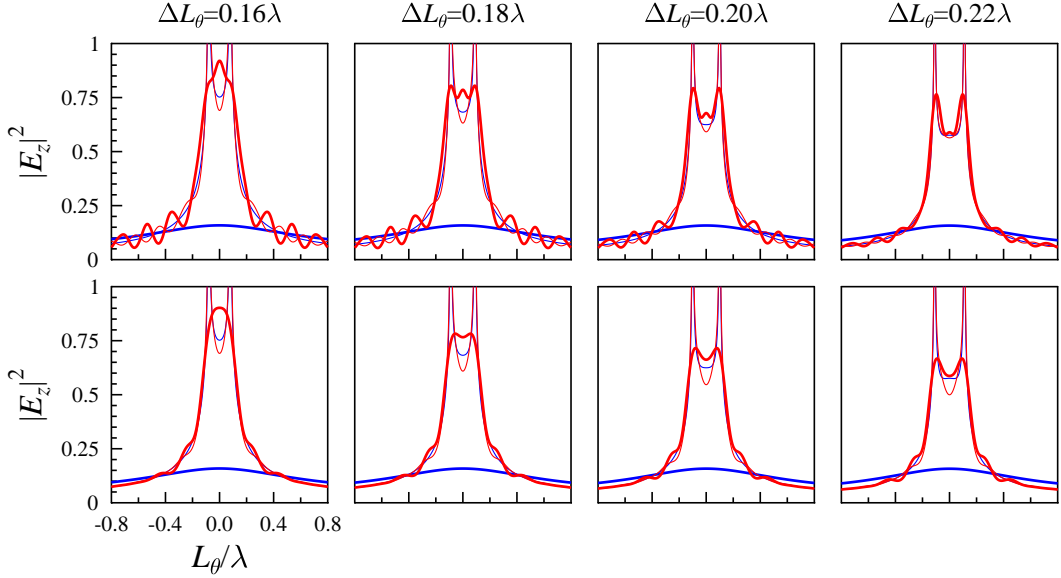}
\caption{Imaging of two parallel line current sources. Panels in
  the first row are for the lens with  $\delta=0.0001-0.00001i$, and those
  in the second row are for the lens with $\delta=0.0001-0.001i$. For
  panels in both rows, from left to right, $\Delta L_\theta$ values
  are respectively $0.16\lambda$, $0.18\lambda$, $0.20\lambda$, and
  $0.22\lambda$. Notice that the image has been stretched by a factor
  of $(2b-a)/a$ due to the higher permittivity inside the lens. Red
  lines: with lens; blue lines: without lens. Thick lines: intensity
  at image plane; thin lines: intensity at object plane.}
\label{fig:src_2}
\end{figure}

With the same two lenses, we numerically demonstrate their inward imaging
of two parallel line current sources. Such numerical experiments provide direct evidence
for the maximum achievable resolution. Note that here we
calculate the distance between two line sources or their images by
considering they are located on a cylindrical plane (concentric to the lens
surfaces). Two $z$-oriented line sources are placed at $r_o=(3b-a)/2$,
and azimuthally separated by a distance denoted as $\Delta
L_\theta$. Their images are recorded while $\Delta L_\theta$ is
decreased gradually. The recorded electric field intensity at object and image
planes for the low-loss and lossy lenses are shown by the
panels in the first and second rows in Fig. \ref{fig:src_2}, respectively.
For the low-loss lens, the intensity at the image plane shows very clearly the presence of
two peaks when $\Delta L_\theta=0.22\lambda$, which corresponds well to
the field at the object plane. As
$\Delta L_\theta$ decreases, the peaks become less evident, and they are
overtaken by a central intensity peak when
$\Delta L_\theta=0.16\lambda$. The central
peak is caused by the extension of a strong surface mode. It is
however arguable that from the image obtained
for $\Delta L_\theta=0.16\lambda$, one still can tell the information of two line
sources. The overall image, though distorted, can possibly be
corrected through signal processing techniques, especially if
one has the preknowledge of the distortion. For the lossy lens, the
information of the two current sources is almost completely lost at
$\Delta L_\theta=0.16\lambda$. However, the two line sources becomes
identifiable once $\Delta L_\theta$ reaches $0.18\lambda$. The
recorded field intensities show much less ripples due to the suppression
of surface modes. In all panels in Fig. \ref{fig:src_2}, the blue
curves represent the field intensities at the object and image planes
while the lens is absent. Drastic decay in amplitude is noticed for
the images recorded without a lens. Hence the restoration of evanescent
EM wave by a cylindrical lens is obvious. Comparing
the low-loss lens to the lossy one, we notice that
the latter suffers less disturbance from surface mode. Due to
this factor, its maximum achievable resolutions for both lenses are
almost comparable.

Lastly, we make a couple of comments on the performance of a cylindrical
lens when some parameter choices for designing the lens are different
from above analyses. First, we comment on the effect of negative slope value used for the
spatial mapping function (Eq. \ref{eq:map}). In practice, one
desires to image an object that is far away from the lens, rather than
maneuvering the lens very close to the object for performing an
imaging. Theoretically this can be achieved by an
increase (negatively) in the negative slope value for the mapping
function. However, for a fixed lens thickness, an increase in the
slope compresses the field in the lens body, or equivalently increases proportionally
the optical lengths for all cylindrical wave components as they
travel across the lens body. That unfortunately amplifies the adverse effect of
material loss to the lens' performance. We noticed that in order to
achieve the same resolution, the thickness of a lens should be
kept inversely proportional to the negative slope,
assuming the same material perturbation. Due to the necessity of
a decrease in lens thickness for maintaining resolution, the object-to-image distance
in fact shortens when the negative slope increases. Another note is that an
increase in the negative slope gives rise
to gradient lens material with higher anisotropic ratio. Therefore
such a way of increasing the object distance outside the lens is not
encouraged. Our second comment is regarding the material perturbation
introduced. So far in our previous theoretical studies we have chosen $\Re(\delta)$ to be positive, while
$\Im(\delta)$ is kept negative. Negative $\Im(\delta)$ is natural for denoting loss incurred by passive
  materials. It is however foreseeable that in practical
  implementations $\Re(\delta)$ can easily be negative. Through
  calculations we have noticed the following situations for choosing
  a negative $\Re(\delta)$. When $|\Re(\delta)|$ is
  relatively larger than $|\Im(\delta)|$, change in the sign of
  $\Re(\delta)$ from positive to negative will affect the
  characteristics of the transfer function only around the resonance
  peak position. Noticeably, the resonance peak will shift slightly to
  a larger $k_\theta$ position, regardless of inward or
  outward operation mode. For example, the resonance peak of the transfer function for the lens with
  $\delta=0.001-0.0001i$ in Fig. \ref{fig:effect_delta} will shift
  from $k_\theta/k_1=4.70$ to $5.05$ if we use
  $\delta=-0.001-0.0001i$. Under the scenario that $|\Re(\delta)|$ is relatively smaller
  than $|\Im(\delta)|$, no significant change in transfer function can
  be noticed when we flip the sign of $\Re(\delta)$.

In conclusion, based on the coordinate transformation technique we
have designed a type of cylindrical superlenses. The transfer function
analyses show that such a cylindrical lens, even with a
wavelength-scale cross-section size, can achieve similar
subwavelength imaging resolution as compared to the previously reported
negative index slab lens. Subwavelength imaging is
demonstrated numerically for cylindrical lenses with imperfect material
parameters. Like its slab lens counterpart, the cylindrical lens
experiences similar limitation as imposed by inevitable material
imperfectness. Most noticeably, at a practical loss level, a
cylindrical lens should have a subwavelength thickness in order to realize
subwavelength focusing. Therefore, these cylindrical lenses are bound
to be ``near-sighted'', a conclusion previously drawn for slab lens
\cite{Podolskiy:OL-2005-lens}. Besides high-resolution imaging, the
proposed cylindrical lens can be applied for lithography
applications. Consider a single E$_z$ polarization operation. A
properly designed cylindrical lens can embed in its interior
a dielectric material with perfect impedance matching. Image can then
be written on the surface of that dielectric material.
The invisibility property of the cylindrical device can also be deployed for non-invasive
electromagnetic probing. Although not
analyzed in this paper explicitly, the coordinate transformation
predicts that the same radial transformation as described in
Eq. \ref{eq:map} will yield a spherical superlens. With the
current fast-developing metamaterial technology, which has already been
used for demonstrating negative refraction \cite{Shelby:NI,zhang:137404}, subwavelength imaging
\cite{Fang:silver,Liu:Science-2007-hyperlens,Smolyaninov:Science-2007-lens},
and even conceptual electromagnetic cloaking \cite{Schurig:cloak}(also
devised in Ref. \cite{Cai:cloakTM} for operating at optical wavelength),
etc., we should be able to realize fabrication of the proposed cylindrical
superlens in near future.

\vspace{0.2in}
\noindent{\bf Acknowledgement:} This work is supported by the Swedish
Foundation for Strategic Research (SSF) through the INGVAR program,
the SSF Strategic Research Center in Photonics, and the Swedish
Research Council (VR).


\begin{thebibliography}{25}
\expandafter\ifx\csname natexlab\endcsname\relax\def\natexlab#1{#1}\fi
\expandafter\ifx\csname bibnamefont\endcsname\relax
  \def\bibnamefont#1{#1}\fi
\expandafter\ifx\csname bibfnamefont\endcsname\relax
  \def\bibfnamefont#1{#1}\fi
\expandafter\ifx\csname citenamefont\endcsname\relax
  \def\citenamefont#1{#1}\fi
\expandafter\ifx\csname url\endcsname\relax
  \def\url#1{\texttt{#1}}\fi
\expandafter\ifx\csname urlprefix\endcsname\relax\def\urlprefix{URL }\fi
\providecommand{\bibinfo}[2]{#2}
\providecommand{\eprint}[2][]{\url{#2}}

\bibitem[{\citenamefont{Leonhardt}(2006)}]{Leonhardt:conformal}
\bibinfo{author}{\bibfnamefont{U.}~\bibnamefont{Leonhardt}},
  \bibinfo{journal}{Science} \textbf{\bibinfo{volume}{312}},
  \bibinfo{pages}{1777} (\bibinfo{year}{2006}).

\bibitem[{\citenamefont{Pendry et~al.}(2006)\citenamefont{Pendry, Schurig, and
  Smith}}]{Pendry:cloak}
\bibinfo{author}{\bibfnamefont{J.~B.} \bibnamefont{Pendry}},
  \bibinfo{author}{\bibfnamefont{D.}~\bibnamefont{Schurig}}, \bibnamefont{and}
  \bibinfo{author}{\bibfnamefont{D.~R.} \bibnamefont{Smith}},
  \bibinfo{journal}{Science} \textbf{\bibinfo{volume}{312}},
  \bibinfo{pages}{1780} (\bibinfo{year}{2006}).

\bibitem[{\citenamefont{Pendry}(2000)}]{Pendry:perfectLens}
\bibinfo{author}{\bibfnamefont{J.~B.} \bibnamefont{Pendry}},
  \bibinfo{journal}{Phys. Rev. Lett.} \textbf{\bibinfo{volume}{85}},
  \bibinfo{pages}{3966} (\bibinfo{year}{2000}).

\bibitem[{\citenamefont{Leonhardt and Philbin}(2006)}]{Ulf:Relativity}
\bibinfo{author}{\bibfnamefont{U.}~\bibnamefont{Leonhardt}} \bibnamefont{and}
  \bibinfo{author}{\bibfnamefont{T.~G.} \bibnamefont{Philbin}},
  \bibinfo{journal}{New J. Phys.} \textbf{\bibinfo{volume}{8}},
  \bibinfo{pages}{247} (\bibinfo{year}{2006}).

\bibitem[{\citenamefont{Schurig et~al.}(2007)\citenamefont{Schurig, Pendry, and
  Smith}}]{Schurig:2007:slabLens}
\bibinfo{author}{\bibfnamefont{D.}~\bibnamefont{Schurig}},
  \bibinfo{author}{\bibfnamefont{J.~B.} \bibnamefont{Pendry}},
  \bibnamefont{and} \bibinfo{author}{\bibfnamefont{D.~R.} \bibnamefont{Smith}},
  \bibinfo{journal}{Opt. Express} \textbf{\bibinfo{volume}{15}},
  \bibinfo{pages}{14772} (\bibinfo{year}{2007}).

\bibitem[{not({\natexlab{a}})}]{note:finiteSize}
\bibinfo{note}{The adverse effect of inevitable truncation in cross-section for
  a slab lens to its imaging performance is studied in L. Chen, S. He, and L.
  Shen, Phys. Rev. Lett. {\bf 92}, 107404 (2004). Note that the cylindrical
  lens proposed in this paper is still infinite in the axial ($z$) direction.
  Therefore truncation in that direction is necessary for implementation. But
  in principle, we can keep the axial size long enough as compared to the
  source size for minimizing the effect of truncation.}

\bibitem[{\citenamefont{Pendry}(2003)}]{Pendry:OE-2003-clens}
\bibinfo{author}{\bibfnamefont{J.~B.} \bibnamefont{Pendry}},
  \bibinfo{journal}{Opt. Express} \textbf{\bibinfo{volume}{11}},
  \bibinfo{pages}{755} (\bibinfo{year}{2003}).

\bibitem[{not({\natexlab{b}})}]{note:eventMap}
\bibinfo{note}{More generally speaking, the cylindrical annular structure
  perfectly duplicates within its enclosed region the EM events happening
  outside the cylinder, or vice versa.}

\bibitem[{\citenamefont{Kildishev and
  Narimanov}(2007)}]{Kildishev:OL-2007-hyperlens}
\bibinfo{author}{\bibfnamefont{A.~V.} \bibnamefont{Kildishev}}
  \bibnamefont{and} \bibinfo{author}{\bibfnamefont{E.~E.}
  \bibnamefont{Narimanov}}, \bibinfo{journal}{Opt. Lett.}
  \textbf{\bibinfo{volume}{32}}, \bibinfo{pages}{3432} (\bibinfo{year}{2007}).

\bibitem[{\citenamefont{Kildishev and
  Shalaev}(2008)}]{Kildishev:OL-2008-engineeringSpace}
\bibinfo{author}{\bibfnamefont{A.~V.} \bibnamefont{Kildishev}}
  \bibnamefont{and} \bibinfo{author}{\bibfnamefont{V.~M.}
  \bibnamefont{Shalaev}}, \bibinfo{journal}{Opt. Lett.}
  \textbf{\bibinfo{volume}{33}}, \bibinfo{pages}{43} (\bibinfo{year}{2008}).

\bibitem[{\citenamefont{Tsang and
  Psaltis}(2008)}]{Tsang:PRB-2008-magnifyingLens}
\bibinfo{author}{\bibfnamefont{M.}~\bibnamefont{Tsang}} \bibnamefont{and}
  \bibinfo{author}{\bibfnamefont{D.}~\bibnamefont{Psaltis}},
  \bibinfo{journal}{Phys. Rev. B} \textbf{\bibinfo{volume}{77}},
  \bibinfo{pages}{035122} (\bibinfo{year}{2008}).

\bibitem[{\citenamefont{Salandrino and
  Engheta}(2006)}]{Salandrini:PRB-2006-hyperlens}
\bibinfo{author}{\bibfnamefont{A.}~\bibnamefont{Salandrino}} \bibnamefont{and}
  \bibinfo{author}{\bibfnamefont{N.}~\bibnamefont{Engheta}},
  \bibinfo{journal}{Phys. Rev. B} \textbf{\bibinfo{volume}{74}},
  \bibinfo{pages}{075103} (\bibinfo{year}{2006}).

\bibitem[{\citenamefont{Jacob et~al.}(2006)\citenamefont{Jacob, Alekseyev, and
  Narimanov}}]{Jacob:OE-2006-hyperlens}
\bibinfo{author}{\bibfnamefont{Z.}~\bibnamefont{Jacob}},
  \bibinfo{author}{\bibfnamefont{L.~V.} \bibnamefont{Alekseyev}},
  \bibnamefont{and}
  \bibinfo{author}{\bibfnamefont{E.}~\bibnamefont{Narimanov}},
  \bibinfo{journal}{Opt. Express} \textbf{\bibinfo{volume}{14}},
  \bibinfo{pages}{8247} (\bibinfo{year}{2006}).

\bibitem[{\citenamefont{Liu et~al.}(2007)\citenamefont{Liu, Lee, Xiong, Sun,
  and Zhang}}]{Liu:Science-2007-hyperlens}
\bibinfo{author}{\bibfnamefont{Z.}~\bibnamefont{Liu}},
  \bibinfo{author}{\bibfnamefont{H.}~\bibnamefont{Lee}},
  \bibinfo{author}{\bibfnamefont{Y.}~\bibnamefont{Xiong}},
  \bibinfo{author}{\bibfnamefont{C.}~\bibnamefont{Sun}}, \bibnamefont{and}
  \bibinfo{author}{\bibfnamefont{X.}~\bibnamefont{Zhang}},
  \bibinfo{journal}{Science} \textbf{\bibinfo{volume}{315}},
  \bibinfo{pages}{1686} (\bibinfo{year}{2007}).

\bibitem[{\citenamefont{Smolyaninov et~al.}(2007)\citenamefont{Smolyaninov,
  Hung, and Davis}}]{Smolyaninov:Science-2007-lens}
\bibinfo{author}{\bibfnamefont{I.~I.} \bibnamefont{Smolyaninov}},
  \bibinfo{author}{\bibfnamefont{Y.-J.} \bibnamefont{Hung}}, \bibnamefont{and}
  \bibinfo{author}{\bibfnamefont{C.~C.} \bibnamefont{Davis}},
  \bibinfo{journal}{Science} \textbf{\bibinfo{volume}{315}},
  \bibinfo{pages}{1699} (\bibinfo{year}{2007}).

\bibitem[{\citenamefont{Rao and
  Ong}(2003{\natexlab{a}})}]{Rao:PRE-2003-transferFunction}
\bibinfo{author}{\bibfnamefont{X.~S.} \bibnamefont{Rao}} \bibnamefont{and}
  \bibinfo{author}{\bibfnamefont{C.~K.} \bibnamefont{Ong}},
  \bibinfo{journal}{Phys. Rev. E} \textbf{\bibinfo{volume}{68}},
  \bibinfo{pages}{067601} (\bibinfo{year}{2003}{\natexlab{a}}).

\bibitem[{\citenamefont{Smith et~al.}(2003)\citenamefont{Smith, Schurig,
  Rosenbluth, Schultz, Ramakrishna, and Pendry}}]{Smith:APL-2003-limitation}
\bibinfo{author}{\bibfnamefont{D.~R.} \bibnamefont{Smith}},
  \bibinfo{author}{\bibfnamefont{D.}~\bibnamefont{Schurig}},
  \bibinfo{author}{\bibfnamefont{M.}~\bibnamefont{Rosenbluth}},
  \bibinfo{author}{\bibfnamefont{S.}~\bibnamefont{Schultz}},
  \bibinfo{author}{\bibfnamefont{S.~A.} \bibnamefont{Ramakrishna}},
  \bibnamefont{and} \bibinfo{author}{\bibfnamefont{J.~B.}
  \bibnamefont{Pendry}}, \bibinfo{journal}{Appl. Phys. Lett.}
  \textbf{\bibinfo{volume}{82}}, \bibinfo{pages}{1506} (\bibinfo{year}{2003}).

\bibitem[{\citenamefont{Podolskiy and
  Narimanov}(2005)}]{Podolskiy:OL-2005-lens}
\bibinfo{author}{\bibfnamefont{V.~A.} \bibnamefont{Podolskiy}}
  \bibnamefont{and} \bibinfo{author}{\bibfnamefont{E.~E.}
  \bibnamefont{Narimanov}}, \bibinfo{journal}{Opt. Lett.}
  \textbf{\bibinfo{volume}{30}}, \bibinfo{pages}{75} (\bibinfo{year}{2005}).

\bibitem[{not({\natexlab{c}})}]{note:aziWV}
\bibinfo{note}{Following the argument, the factor $\mbox{exp}(im\theta)$ in
  Eqs. \ref{eq:ez1}-\ref{eq:ez3} can be re-configured as
  $\mbox{exp}(ik_\theta(m, r)L_\theta(r))$, where $L_\theta=\theta r$ is the
  azimuthal distance.}

\bibitem[{\citenamefont{Rao and
  Ong}(2003{\natexlab{b}})}]{Rao:PRE-2003-Amplification}
\bibinfo{author}{\bibfnamefont{X.~S.} \bibnamefont{Rao}} \bibnamefont{and}
  \bibinfo{author}{\bibfnamefont{C.~K.} \bibnamefont{Ong}},
  \bibinfo{journal}{Phys. Rev. B} \textbf{\bibinfo{volume}{68}},
  \bibinfo{pages}{113103} (\bibinfo{year}{2003}{\natexlab{b}}).

\bibitem[{\citenamefont{Shelby et~al.}(2001)\citenamefont{Shelby, Smith, and
  Schultz}}]{Shelby:NI}
\bibinfo{author}{\bibfnamefont{R.~A.} \bibnamefont{Shelby}},
  \bibinfo{author}{\bibfnamefont{D.~R.} \bibnamefont{Smith}}, \bibnamefont{and}
  \bibinfo{author}{\bibfnamefont{S.}~\bibnamefont{Schultz}},
  \bibinfo{journal}{Science} \textbf{\bibinfo{volume}{292}},
  \bibinfo{pages}{77} (\bibinfo{year}{2001}).

\bibitem[{\citenamefont{Zhang et~al.}(2005)\citenamefont{Zhang, Fan, Panoiu,
  Malloy, Osgood, and Brueck}}]{zhang:137404}
\bibinfo{author}{\bibfnamefont{S.}~\bibnamefont{Zhang}},
  \bibinfo{author}{\bibfnamefont{W.}~\bibnamefont{Fan}},
  \bibinfo{author}{\bibfnamefont{N.~C.} \bibnamefont{Panoiu}},
  \bibinfo{author}{\bibfnamefont{K.~J.} \bibnamefont{Malloy}},
  \bibinfo{author}{\bibfnamefont{R.~M.} \bibnamefont{Osgood}},
  \bibnamefont{and} \bibinfo{author}{\bibfnamefont{S.~R.~J.}
  \bibnamefont{Brueck}}, \bibinfo{journal}{Phys. Rev. Lett.}
  \textbf{\bibinfo{volume}{95}}, \bibinfo{pages}{137404}
  (\bibinfo{year}{2005}).

\bibitem[{\citenamefont{Fang et~al.}(2005)\citenamefont{Fang, Lee, Sun, and
  Zhang}}]{Fang:silver}
\bibinfo{author}{\bibfnamefont{N.}~\bibnamefont{Fang}},
  \bibinfo{author}{\bibfnamefont{H.}~\bibnamefont{Lee}},
  \bibinfo{author}{\bibfnamefont{C.}~\bibnamefont{Sun}}, \bibnamefont{and}
  \bibinfo{author}{\bibfnamefont{X.}~\bibnamefont{Zhang}},
  \bibinfo{journal}{Science} \textbf{\bibinfo{volume}{308}},
  \bibinfo{pages}{534} (\bibinfo{year}{2005}).

\bibitem[{\citenamefont{Schurig et~al.}(2006)\citenamefont{Schurig, Mock,
  Justice, Cummer, Pendry, Starr, and Smith}}]{Schurig:cloak}
\bibinfo{author}{\bibfnamefont{D.}~\bibnamefont{Schurig}},
  \bibinfo{author}{\bibfnamefont{J.~J.} \bibnamefont{Mock}},
  \bibinfo{author}{\bibfnamefont{B.~J.} \bibnamefont{Justice}},
  \bibinfo{author}{\bibfnamefont{S.~A.} \bibnamefont{Cummer}},
  \bibinfo{author}{\bibfnamefont{J.~B.} \bibnamefont{Pendry}},
  \bibinfo{author}{\bibfnamefont{A.~F.} \bibnamefont{Starr}}, \bibnamefont{and}
  \bibinfo{author}{\bibfnamefont{D.~R.} \bibnamefont{Smith}},
  \bibinfo{journal}{Science} \textbf{\bibinfo{volume}{314}},
  \bibinfo{pages}{977} (\bibinfo{year}{2006}).

\bibitem[{\citenamefont{Cai et~al.}(2007)\citenamefont{Cai, Chettiar,
  Kildishev, and Shalaev}}]{Cai:cloakTM}
\bibinfo{author}{\bibfnamefont{W.}~\bibnamefont{Cai}},
  \bibinfo{author}{\bibfnamefont{U.~K.} \bibnamefont{Chettiar}},
  \bibinfo{author}{\bibfnamefont{A.~V.} \bibnamefont{Kildishev}},
  \bibnamefont{and} \bibinfo{author}{\bibfnamefont{V.~M.}
  \bibnamefont{Shalaev}}, \bibinfo{journal}{Nat. Photonics}
  \textbf{\bibinfo{volume}{1}}, \bibinfo{pages}{224} (\bibinfo{year}{2007}).

\end{thebibliography}
\end{document}